\documentclass[11pt, draftclsnofoot, onecolumn]{IEEEtran}

\usepackage{subfig,cite,graphicx,amsmath,amssymb}
\usepackage{fancyhdr}
\usepackage{mdwmath}
\usepackage{mdwtab}
\usepackage{balance}
\usepackage{xcolor}
\usepackage{bm}
\usepackage{amsthm}
\usepackage{algorithm}
\usepackage{algorithmic}
\usepackage{multirow}
\usepackage{flafter}
\usepackage{siunitx}
\usepackage{url}
\usepackage{epsfig}
\usepackage{hyperref}

\usepackage{footnote}
\usepackage{verbatim}
\usepackage{mathrsfs}
\usepackage{lipsum}
\usepackage{mathrsfs}%
\usepackage{blindtext}
\usepackage{colortbl}
\makeatletter
\def\BState{\State\hskip-\ALG@thistlm}
\makeatother
\usepackage{cite}
\usepackage[T1]{fontenc}
\IEEEoverridecommandlockouts

\begin{document}
\title{Deep Learning in Physical Layer Communications}

\author{
\IEEEauthorblockN{Zhijin~Qin,~
Hao Ye,~
Geoffrey Ye Li, and Biing-Hwang Fred Juang
}
 \thanks{Zhijin Qin is with Queen Mary University of London, London E1 4NS, U.K., (email: z.qin@qmul.ac.uk).}
\thanks{Hao Ye, Geoffrey Ye Li, and Biing-Hwang Fred Juang are with Georgia Institute of Technology, Atlanta, GA 30332 USA, (email: yehao@gatech.edu; liye@ece.gatech.edu, juang@ece.gatech.edu).}
}

\maketitle

\begin{abstract}
Deep learning (DL) has shown great potentials to revolutionizing communication systems. This article provides an overview on the recent advancements in DL-based  physical layer communications. DL can improve the performance of each individual block in communication systems or optimize the whole transmitter/receiver. Therefore, we categorize the applications of DL in physical layer communications into  systems with and without block  structures. For the DL-based communication systems with the block structure, we demonstrate the power of DL in signal compression and signal detection. We also discuss the recent endeavors in developing DL-based end-to-end  communication systems. Finally, the potential research directions are identified to boost the intelligent physical layer communications.
\end{abstract}


\section{Introduction}
 The idea of using neural networks (NN) to intelligentize machines can be traced to 1942 when a simple model was proposed to simulate the status of a single neuron. Deep learning (DL) adopts a deep neural network (DNN) to find data representation at each layer, which could be  built by using different types of   machine learning (ML) techniques, including supervised ML, unsupervised ML, and reinforcement learning. In recent years, DL has shown its overwhelming privilege in many areas, such as computer vision, robotics, and natural language processing, due to its advanced algorithms and tools in learning complicated models.

Different from the aforementioned DL applications, where it is normally difficult to find a concrete mathematical model for feature representation, various theories and models, from information theory to channel modelling, have been well developed to describe communication systems~\cite{physical_layer}. However, the gap between theory and  practice motivates us to work on intelligent communications. Particularly, the following challenges have been identified in the existing physical layer communications:
\begin{itemize}
  \item Mathematical model versus practical  imperfection: The conventional communication systems rely on the mathematically expressed  models for each block. While in the real-world applications, complex systems may contain unknown effects that are difficult to be expressed analytically. For example, it is  hard to model underwater acoustic channels or molecular communications. Therefore, a more adaptive framework is required to handle the challenges.
  \item Block structures versus global optimality: The traditional communication systems consist of several processing blocks, such as channel encoding, modulation, and signal detection, which are designed and optimized within each block locally. Thus  the global optimality cannot be guaranteed. Moreover, the optimal communication system structure varies with  environments. As a result,  optimal or robust communication systems for different scenarios are more than desired.

\end{itemize}

DL could be a pure data-driven method, where the networks/systems are optimized over a large training data set and  a mathematically tractable model is unnecessary. Such a feature motivates us to exploit DL in communication systems in order to address the aforementioned challenges.  In this situation, communication systems can be optimized for  specific hardware configuration and channel to address the imperfection issues. On the other hand, many models in physical layer communications have been established by researchers and engineers during the past several decades. Those models can be combined with DL to design model-driven DL-based communication systems, which can take advantages of both model-based algorithms and DL~\cite{He:ModelSurvey}.

There is evidence that the ``learned'' algorithms could be executed faster with lower power consumption than the existing manually ``programmed'' counterparts as NNs can be highly parallelized on the concurrent architectures and implemented with low-precision data types. Moreover, the passion on developing artificial intelligence-powered devices from  manufacturers, such as Intel$^{\tiny{\copyright}}$ Movidius$^\text{TM}$ Neural Compute Stick, has also boosted the boom of DL-based wireless communications.

This article will identify  the gains that DL can bring to wireless physical layer communications, including  the  systems with the block structure and the end-to-end structure merging  those blocks. The rest of this article is organized as follows. Section~\ref{s} introduces the important basis of DNN and illustrates  DL-based communication systems. Section~\ref{s1} discusses how to apply DL to block-structured communication systems. Section~\ref{s4} demonstrates DL-based end-to-end communication systems, where individual block for a specific function, such as channel estimation or decoding, disappears. Section~\ref{s5} concludes this article with potential research directions in the area of DL-based physical layer communications.

\section{Deep Neural Networks and Deep Learning Based Communications}\label{s}
In this section, we will first introduce the basis of DNN, generative adversarial network (GAN), conditional GAN, and Bayesian optimal estimator, which are widely used in DL-based communication systems. Then we will discuss the intelligent communication systems with DL.

\subsection{Deep Neural Networks}
\subsubsection{Deep Neural Networks Basis}
\begin{figure}[!t]
\centering
\includegraphics[width=6.2in]{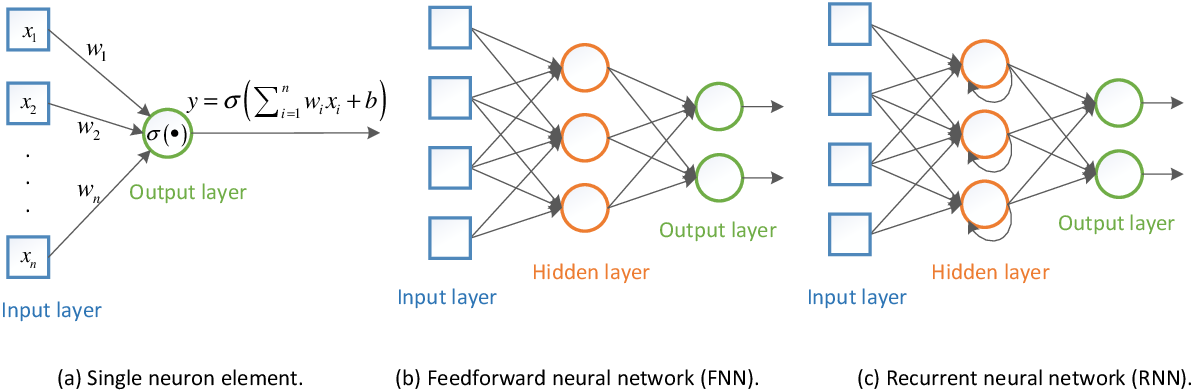}
\caption{Development of neural networks.}
\label{DNN}
\end{figure}

As aforementioned, research on NN started from the single neuron.  As shown in Fig.~\ref{DNN} (a), the inputs of the NN are  $ \left\{ {{x_1},{x_2}, \ldots ,{x_n}}\right\} $ with the corresponding  weights, $\left\{{w_1},{w_2}, \ldots ,{w_n}\right\}$. The neuron can be represented by a non-linear activation function, $\sigma \left(  \bullet  \right)$, that takes the sum of the weighted inputs. The output of the neuron can be expressed as $y = \sigma \left( {\sum\nolimits_{i = 1}^n {{w_i}{x_i}} } +b \right)$, where $b$ is the shift of the neuron. 
An NN can be established by connecting multiple neuron elements to generate multiple outputs to construct a layered architecture. In the training process, the labelled data, i.e., a set of input and output vector pairs, is used to adjust the weight set, $\mathcal{W}$, by  minimizing a loss function. In the NN with single neuron element, $\mathcal{W}=\left\{ {b,{w_1},{w_2}, \ldots ,{w_n}} \right\}$. The commonly-used loss functions include  mean-squared error (MSE) and categorical cross-entropy. To train the model for a specific scenario,  the loss function can be revised by introducing the $l_1$- or $l_2$-norm of $\mathcal{W}$ or activations. $l_1$- or $l_2$-norm of $\mathcal{W}$ can also introduced in the loss function as the regularizer to improve the generalization capabilities. Stochastic gradient descent (SGD) is one of the most popular algorithms to optimize $\mathcal{W}$.

With the layered architecture, a DNN includes multiple fully connected hidden layers, in which each of them represents a different feature of the input data. Fig.~\ref{DNN} (b) and (c) show two typical DNN models:  feedforward neural network (FNN) and recurrent neural network (RNN). In FNNs, each neuron is connected to the adjacent layers while the neurons in the same layers are not connected to each other. The deep convolutional network (DCN) is developed from the fully connected FNN by only keeping some of the connections between neurons and their adjacent layers. As a result, DCN  can significantly reduce the number of  parameters to be trained~\cite{Rafael:SPM:2018}. Recently, DL has boosted many applications due to the powerful algorithms and tools.  DCN has shown its great potential for signal compression and recovery problems, which will be demonstrated in Section~\ref{s2}.

For the RNN in Fig.~\ref{DNN} (c), the outputs of each layer are determined by both the current inputs and their hidden states in the previous time step. The critical difference between FNN and RNN is that the latter has memory and can capture the  hidden layer outputs in the previous step. However, as RNN  is dependent on time over a long term, non-stationary errors may show up during the training process. A special type of RNN, named long short-term memory (LSTM), has been further proposed to eliminate some unnecessary information in the network. LSTM has been widely applied in various cases, such as the joint deign of source-channel coding, which will be briefly discussed in Section~\ref{s2}.

\subsubsection{Generative Adversarial Net (GAN)  and Conditional GAN}
\begin{figure}[!t]
\centering
\includegraphics[width=0.75\linewidth]{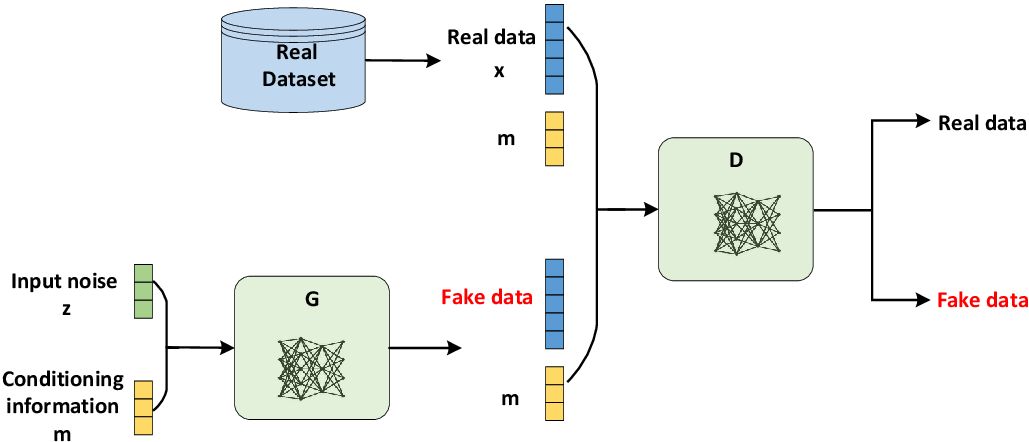}
\caption{Structure of conditional GAN~\cite{Hao_GAN}.} \label{fig:CGAN}
\end{figure}

Training a typical DNN is heavily dependent on the large amount of labelled data, which may be difficult to obtain or even unavailable in certain circumstance. As shown in Fig.~\ref{fig:CGAN}, GAN is a type of generative method, which can produce data that follows certain target distribution. By doing so, demands for the amount of labelled data can be lowered. In Fig.~\ref{fig:CGAN}, a GAN consists of a generator, $G$,  and a discriminator, $D$. $D$ attempts to differentiate between the real data and the fake data generated by  $G$ while $G$ tries to generate plausible data to fool $D$ into making mistakes, which introduces min-max two player game between $G$ and  $D$. As result of the min-max two player game, the generator, $G$, will generate  data with the same distribution as the real data and therefore, the discriminator, $D$, cannot identify the difference between the real and fake data. Conditional GAN is an extension of GAN by providing extra conditioning information,  $\mathbf{m}$, where the conditioning information has been fed to both $G$ and $D$  as the additional input.

In communication systems, a GAN and a conditional GAN can be applied to model the distribution of the channel output. Moreover, the learned model can be utilized as a surrogate of the real channel when training the transmitter so that the gradients can pass through to the transmitter. An application example of conditional GAN will be introduced in Section~\ref{GAN}.

\subsection{Bayesian Optimal Estimator}

Besides the standard DL models,  the expert knowledge can be beneficial in modifying the structures of  DL models to provide more explainable and predictable models in physical layer communications. In fact, many signal processing modules in communication systems, such as the multi-input multi-output (MIMO) detection and channel decoding,  can be cast as posterior probability inference problems in probabilistic graphical models, where the dependence of observation variables (e.g. the received signals) and the latent variables (e.g. the transmitted signals) are expressed explicitly. The posterior distribution of the latent variables can be calculated via Bayesian inference and then the Bayesian optimal estimators are obtained by minimizing the expected MSE with given posterior distributions.

Even though  the exact computation of the Bayesian optimal estimators are computationally intractable in many problems, some iterative approaches, such as the approximate message passing (AMP) and expectation-propagation, can approximate the performance of Bayesian optimal efficiently. Moreover, these iterative detectors can be further improved by unfolding and representing the iterative procedures with DL models, where the parameters of the model can be updated based on the training data. A detailed example will be shown in Section~\ref{bayes}.

\subsection{Deep Learning based Communications}
\begin{figure}[!t]
\centering
\includegraphics[width=3.5in]{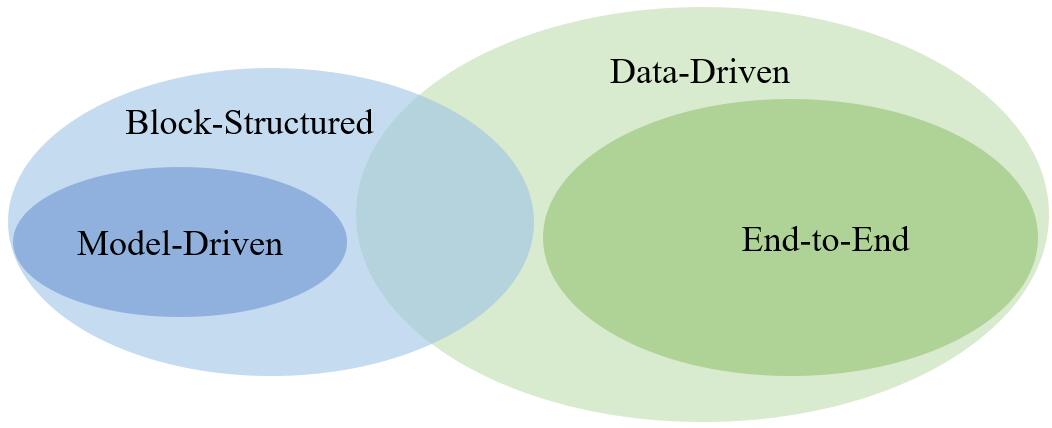}
\caption{Intelligent communication system structure.}
\label{new_structure}
\end{figure}

Fig.~\ref{new_structure} shows the intelligent communication system structure with DL. Compared to the conventional block-based communication structure, DL-based communication systems utilize the power of DL to facilitate transmission. A data-driven DL-based communication system is usually represented by a DNN and  a large amount of labelled data is used to tune the parameters of the DNN. Such a DNN can be regarded as a `black box' and  used in each processing block individually to replace the existing algorithms, which is the overlap of block-structured and data-driven modules shown in Fig.~\ref{new_structure}. Moreover, it  can be used to optimize the whole transmitter or the whole receiver when a DNN merges all processing blocks in the communication system. Such a structure is named as an end-to-end communication system. Examples of data-driven DL-based  communication systems will be introduced in Section~\ref{s1} and Section~\ref{s4}.

Typically, training a fully connected DNN requires plenty of training time in addition to a huge data set,  especially in end-to-end communication systems. However, the computing resources and labelled data are often scarce in wireless communication systems. As shown in Fig.~\ref{new_structure}, model-driven DL methods exploit  the known physical mechanism and domain knowledge, such as well-developed channel models and information theory, which can reduce the number of parameters to be learned and improve the training efficiency of some block-structured communication systems. An example of model-driven DL based wireless communications will be provided in Section~\ref{bayes}.

The model-driven methods exploit some prior knowledge of system to reduce the number of parameters to be learned. While the data-driven methods assume a general system structure that usually has lots of unknown parameters to be trained by a huge data set. Each of them has its advantages and disadvantages. In general, it involves the  trade-off between variance and bias in the learning theory. With prior knowledge, the sample complexity for learning models can be largely reduced, but the models may suffer when the prior knowledge is not accurate in the real scenario. On the contrary, the data-driven model is  with less presumption. The sample complexity is large but it can be more robust under variant circumstances.

\section{Deep Learning Based Block-Structured Communications}\label{s1}
Even though the existing block-structured  communication systems have been carefully designed  from their infancy to the fifth generation (5G), more efforts are still required to break the bottleneck in wireless communication systems. In this section, we focus on the applications of DL in different communication blocks, which are categorized into intelligent signal compression and  detection.

\subsection{Intelligent Signal Compression}\label{s2}
Most types of source data exhibit  unique inner structures that can be utilized for  compression. Such structured data can be modelled by different approaches. Sparse representation is a commonly-used one. It is worth noting that the most important property of DL is that it can automatically find compact low-dimensional represeations/featues of high dimensional data~\cite{KAI:SPM:2017,QIN:SPM:2018}, which can be demonstrated by the following two examples.

\subsubsection{Model-Driven CSI Feedback Compression and Reconstruction}
In the downlink of frequency division duplex networks, massive MIMO relies on channel state information (CSI) feedback to achieve performance gains from multiple antennas at the base station. However, the large number of antennas results in excessive feedback overhead. Extensive work has been carried out to reduce the feedback overhead by utilizing the spatial and temporal correlations of CSI. By exploiting the sparse property of CSI, compressive sensing (CS) has been applied to compress  CSI at the user side and  the compressed CSI is then recovered at the base station. However,  traditional CS algorithms face challenges as real-world data is not exactly sparse and the convergence speed of the existing signal recovery algorithms is relatively slow, which has limited the practical applications of CS~\cite{DeepInverse:2017}.

DCN has been applied to learn the inverse transformation from measurement vectors to signals to improve the recovery speed in CS~\cite{DeepInverse:2017}. Particularly, DCN has two distinctive features that make it uniquely applicable to sparse recovery problems. One is that the neurons are  sparsely connected. The other is  with shared weights across the entire receptive fields of one layer. DCN can increase the learning speed  comparing to a fully-connected network~\cite{Papyan:SPM:2018}. Learned denoising-based AMP (LDAMP)~\cite{Baraniuk:NIPS:2017} is one of the excellent signal recovery algorithms  in terms of both accuracy and speed, which has been applied to channel estimation in millimeter-wave (mmWave) communications~\cite{He:ModelSurvey}. However, the achieved improvement still cannot boost the CS-based CSI estimation.

\begin{figure}[!t]
\centering
\includegraphics[width=6.0in]{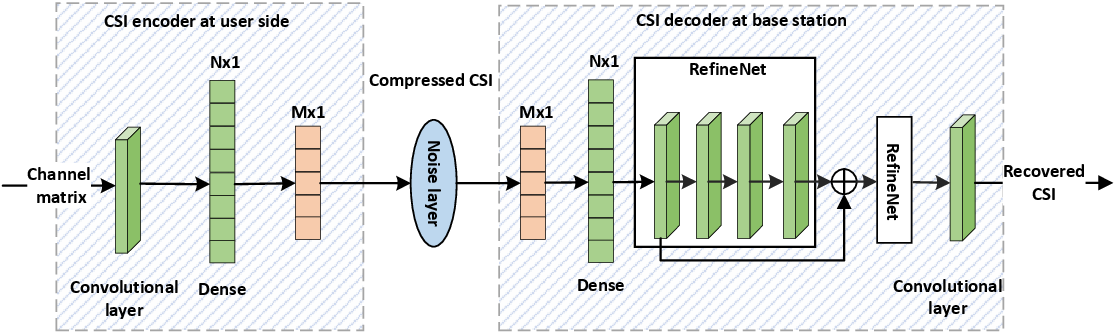}
\caption{DL-based channel compression, feedback, and recovery by CsiNet~\cite{Wen:WCL:2018}.}
\label{block_system}
\end{figure}

CsiNet~\cite{Wen:WCL:2018} has been proposed  to mimic the CS  processes for channel compression, feedback, and  reconstruction. Particularly, an encoder is used to collect compressed measurements  by directly learning channel structures from the training data. As shown in Fig.~\ref{block_system}, by taking the channel matrix in the angular-delay domain as  inputs, the first layer of the encoder is a convolutional layer that generates two feature maps. Then the feature maps are vectorized and a fully connected layer is used to generate the real-valued compressed CSI. Only those compressed CSI is fed back to the base station. With such an encoder, the feedback overhead is significantly reduced. At the base station,  the decoder reconstructs the CSI  via learning the inverse transformation of the compressed feedback. It has been shown that CsiNet remarkably outperforms the traditional CS-based methods in terms of both compression ratio and recovery speed.

\subsubsection{Data-Driven Joint Source-Channel Coding}
The typical source coding mainly utilizes the sparse property  to remove the redundancy in  source data while channel coding  improves the robustness to noise by adding redundancy to the coded information when it is transmitted over channels. Shannon separation theorem guarantees that source coding and channel coding can be designed separately without loss of optimality. However, in many communication systems,  source coding and channel coding are designed jointly as it is not practical to have very large blocks.

A joint source-channel coding based on DL has been proposed in~\cite{GDSmith}. With text as the source data, the DL-based source-channel encoder and decoder may output different sentences but preserving their semantic information content. Specifically, the proposed model adopts a RNN encoder, a binarization layer, a channel layer, and a RNN decoder. The text is structured before it is processed by the stacked bidirectional LSTM  networks. Then the binarizer is adopted to output binary values, which are taken as the inputs of the channel. At the receiver, a stack of LSTM is used for decoding. By doing so, the word-error rate is lowered compared with various traditional separate source-channel coding baselines, such as using huffman and Reed-Solomon code for source and channel coding, respectively. Even though this design is particularly for text processing, it inspires us to apply DL to  where recovery of the exact transmitted data is not compulsory as long as the main information within it is conveyed. For example, in sparse support detection, we need to determine if there is a sparse support at each location while the exact amplitude of each location is not of interest.

In addition to the aforementioned two examples, DL has also been widely applied in other signal compression applications. For example, instead of performing joint source-channel coding, DL can be applied to source coding and channel coding, separately, to achieve better performance compared to typical coding techniques. Moreover, DNN has also been widely applied to facilitate the design of measurement matrix and signal recovery algorithm in CS~\cite{DeepInverse:2017}, which can be used in various wireless scenarios, i.e., channel estimation and wideband spectrum sensing.

\subsection{Intelligent Signal Detection}\label{bayes}

The DL-based detection algorithms can significantly improve the performance of communication systems, especially when the joint optimization of the traditional processing blocks is required and when the channels are hard to be characterized by analytical models. Here, we provide two examples for DL-based detection.

\subsubsection{Data-Driven Joint Channel Estimation and Signal Detection}
\begin{figure}[!t]
\centering
\includegraphics[width=0.6\linewidth]{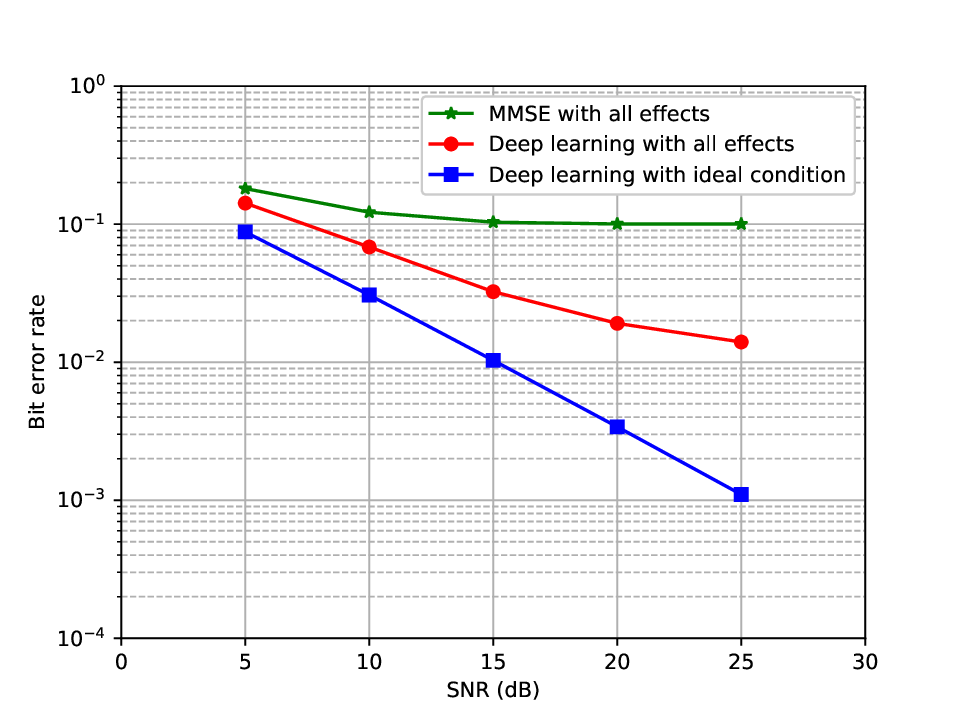}
\caption{Performance comparison of DL and minimum MSE-based joint channel estimation and signal detection in OFDM systems~\cite{Hao}.}
\label{fig:all_effect}
\end{figure}

Traditionally,  channel estimation and signal detection are two separate procedures at the receiver. The CSI is first estimated by means of pilots prior to the detection of the transmit symbols. Then with the estimated CSI, the transmit symbols can be recovered at the receiver. A joint channel estimation and signal detection approach has been proposed in~\cite{Hao}. Specifically, a five-layer fully connected DNN is embedded into an orthogonal frequency-division multiplexing (OFDM) receiver for joint channel estimation and detection by treating the channel as a  `black box'.  The DNN is trained to reconstruct the transmit data by feeding the received signals corresponding to the transmit data and pilots as inputs. Therefore, the channel information can be inferred implicitly by the DNN and used to predict the transmit data directly without explicitly estimating the CSI. Fig.~\ref{fig:all_effect} demonstrates that the DNN-based channel estimation and detection method outperforms the minimum MSE-based approach when without adequate pilots or cyclic prefix, and with nonlinear distortion\footnote{The data set and simulation codes can be downloaded from \url{https://github.com/haoyye/OFDM_DNN}.}. The advantage is that when these adversarial effects occur, the data-driven method can learn to deal with these effects in a supervised manner, i.e., updating the parameters to minimize the empirical loss, which improves the robustness to the undesired circumstances.

\subsubsection{Model-Driven MIMO Detection}
In MIMO detection, iterative methods, which are based on Bayesian optimal detectors, have shown superior performance with moderate computation complexity. However, these detectors often impose assumptions on the channel distribution, which limits the performance under many complicated environments. By incorporating learning based approaches, the adaptability of the detectors can be improved since the parameters of the model can be refined according to the specific data.  In ~\cite{He:ModelSurvey}, the iterative procedures are unfolded to a signal flow graph.
Only several critical variables are required to tune the graph in the supervised learning manner. This trainable framework has been combined with the orthogonal AMP detector, where only two variables are set as the trainable parameters in each iteration. Since the number of trainable parameters are comparable to that of iterations, it can be easily trained within a shorter period and with less training data comparing to a regular DNN  while improving the performance of the orthogonal AMP detector in Rayleigh and correlated MIMO channels. Therefore, this approach can be scaled to massive MIMO communications with great potentials to be applied  to  time-varying channels.

Apart from the wireless signal compression and  detection, DL has been exploited for various tasks in  physical layer communications. Compared with the traditional methods, it has shown higher robustness to channels. For example, DNN has been utilized in the channel decoding and is more robust to variations of the additive white Gaussian noise (AWGN) channel model \cite{Kim:ICLR:2018}. In addition, DL can improve the system performance by exploiting the additional contextual information. For example, in mmWave systems, DL can be used for beam prediction, where some contextual information, such as the locations of the receiver and the surrounding vehicles in vehicular networks, can be taken into consideration to improve the prediction. Moreover, DL has shown its privilege on the molecular signal detection when the channel models are optimized based on training data instead of any prior channel information.

\section{Deep Learning Based End-to-End Communications}\label{s4}
In the previous section, we have discussed the applications of  DL in each individual block of communication systems. In this section, we will present innovative learning-based communication systems by treating the entire communication system as an end-to-end reconstruction task~\cite{Dorner:JSTSP:2018,RL_E2E,Hao_GAN}. Particularly, based on the data-driven methods, the transmitter learns to encode the source data into encoded symbols (or transmit signals) to be transmitted over the channel while the receiver learns to recover the source data from the received signals. The weights of the model are optimized in a supervised learning manner based on an end-to-end loss on the recovery accuracy. By doing so, the block structure in the conventional communication systems is no longer required. Moreover, the end-to-end method has great potentials to provide a universal solution for different channels.

\begin{figure}[!t]
\centering

\subfloat[Reinforcement learning based end-to-end communication systems.]{
	\label{Reinforce_e2e}
	\includegraphics[width=0.85\textwidth]{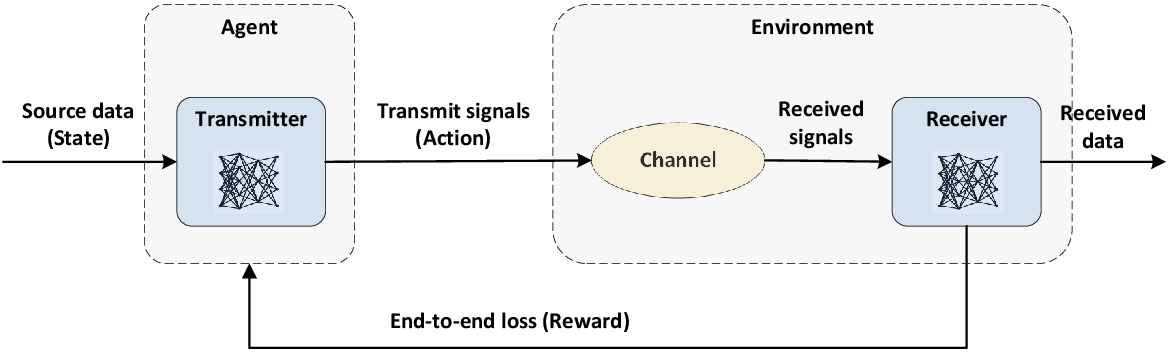} }

\subfloat[Conditional GAN based end-to-end communication systems.]{
	\label{CGAN_e2e}
	\includegraphics[width=0.85\textwidth]{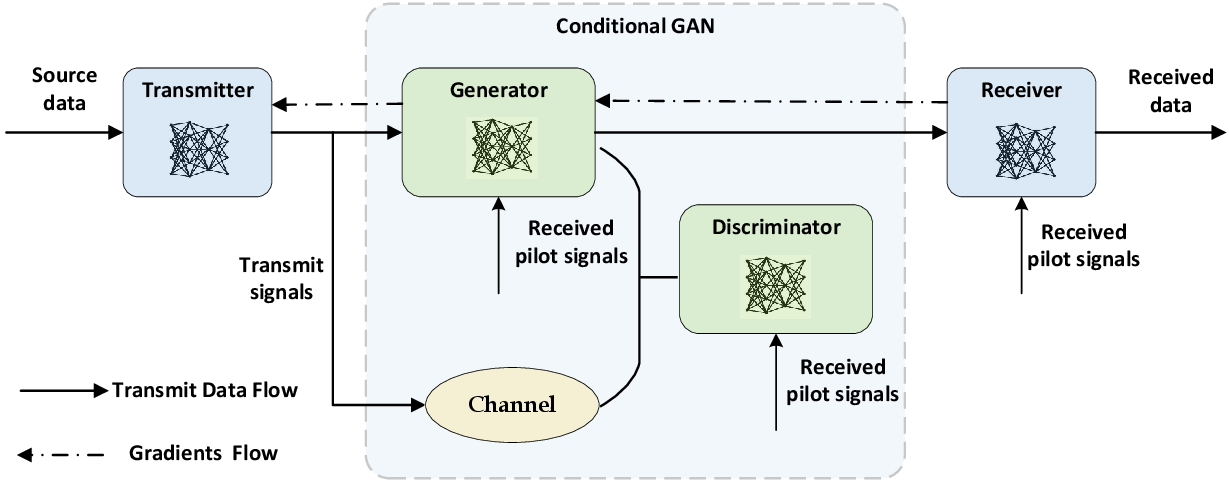} }
\caption{End-to-end communication system models.}
\label{end-end}
\end{figure}

As aforementioned, the weights of the DNN are usually learned based on the SGD with the gradients of the loss function back-propagated from the output layer to the input layer. Nevertheless, when the channel parameters are unknown in advance, the gradients cannot back-propagate through the unknown channel since the gradients for updating the transmitter is blocked by the unknown channel, which forestalls the learning of the end-to-end networks. The channel transfer function may be pre-assumed to solve the issue, but any such assumption would bias the learned models, repeating the pitfalls resulted from the likely discrepancy between the assumed channel models and the actual channels. In addition, in real communication systems, an accurate channel transfer function is difficult to obtain in advance since the end-to-end channel often embraces different types of random effects, such as channel noise and time-varying, which may be unknown or cannot be expressed analytically. As shown as in Fig.~\ref{end-end}, we will introduce two methods to address the issue in this section.

\subsection{Reinforcement Learning Based End-to-End Systems}
In \cite{RL_E2E}, a reinforcement learning based approach has been proposed to circumvent the problem of missing gradients from  channels when optimizing the transmitter. As shown in Fig. \ref{end-end} (a), the transmitter, converting the source data into the transmit symbols, is considered as an agent while both the channel and the receiver are regarded as the environment. The agent will learn to take  actions to maximize the cumulative rewards emitted from the environment. At each time, the transmit data is regarded as the state observed by the transmitter and the transmit signals are regarded as the action taken by the transmitter. The end-to-end loss on each sample will be calculated at the receiver and fed back to the transmitter as the reward from the environment, which guilds the training of the transmitter. By using the policy gradient algorithm, a standard reinforcement learning approach, the transmitter can learn to maximize the reward, i.e., optimize the end-to-end loss, without requiring the gradients from the channel.

\subsection{Conditional GAN Based End-to-End Systems}\label{GAN}

In order to solve the  missing gradient problem  and lower the demands for the large amount of training data, a generative approach based on conditional GAN has been proposed in \cite{Hao_GAN}. As in Fig. \ref{end-end} (b), the end-to-end learning of a communication system is enabled without requiring prior information of the channel by modelling the conditional distribution of the channel.
In Fig. \ref{end-end} (b), the end-to-end pipeline consists of DNNs for the transmitter, the channel generator, and the receiver, which are trained iteratively.
Since the conditional GAN learns to mimic the channel effects, it acts as a surrogate channel for the gradients to pass through, which enables the training of the transmitter. The conditioning information for the conditional GAN is the transmit signals from the transmitter along with the received pilot information used for estimating the channel. Therefore, the generated output distribution will be specific to the instantaneous channel and the transmit signals. As a result, the conditional GAN based end-to-end communication system can be applied to more realistic time-varying channels. The simulation results in~\cite{Hao_GAN} confirm the effectiveness of the conditional GAN based end-to-end communication system, by showing similar performance as the Hamming (7,4) code with maximum-likelihood decoding (MLD).

\section{Conclusions and Future Directions}\label{s5}
We has demonstrated great potentials  of DL in physical layer communications in the above. By summarizing how to apply DL in communication systems,  the following research directions have been identified to bring the intelligent physical layer communications from theory to practice.

\subsection{Can DL-based End-to-End Communications Beat the Traditional?}
We have briefly introduced end-to-end communications in Section~\ref{s4}. From the initial research results in~\cite{Hao_GAN} and~\cite{RL_E2E}, the performance of DL-based end-to-end communications is comparable with the traditional ones. However, it is not clear whether the DL-based end-to-end communications eventually outperform the traditional ones in terms of performance and complexity or how much gain can be achieved. We are expecting the answers to these questions soon.

\subsection{Tradeoff between System Performance and Training Efficiency}
The existing work has shown the power of data-driven models in physical layer communications. Even though a universal transmitter/receiver can be optimized in  the end-to-end learning-based communication design, the training process takes very long  as all the communication blocks are merged. In order to improve the training efficiency and achieve good system performance, part of the communication blocks can be kept and model-driven DL methods can be considered. Then we need to carefully design the system to achieve a good tradeoff between the training efficiency and system performance.

\subsection{Communication Metric Learning}
In traditional communication systems, the objective is the error-free reconstruction of the transmit data. While in real applications, the objective of sharing the information may vary from task to task and the reconstruction metrics may not be satisfactory for all tasks. For instance, bit-error rate is not a good metric for images and videos transmission since it cannot reflect the properties of human visual perception. In the end-to-end communication systems, the metric should be revised to address  specific requirements for each application. The basic idea is that the transmit data will not be treated as equally important, the recovered data may contain transmission error, but the semantic information contained in the data, which is further employed for the application-specific tasks at the receiver, should remain intact.

\subsection{Open Access Real-World Data Sets}
The bloom of various applications of learning techniques  should be largely credited to the accessible real-world data sets, such as ImageNet for computer vision. These open access data sets provide an efficient way to compare the performance of different learning algorithms. However, such a type of accessible data sets for wireless communications are still under developed. The data protection and privacy regulations further limit the open access of real-world communication data. However, it is still essential to publish some data sets, i.e., channel responses under different channel conditions, to speed up the development of DL-based physical layer communications.

\bibliographystyle{IEEEtran}

\end{document}